# Superstatistics from a different perspective

## F. Sattin

*Consorzio RFX*

*Corso Stati Uniti 4, 35127 Padova (Italy)*

*e-mail: fabio.sattin@igi.cnr.it*

**Abstract**

*In this paper we elaborate on the recently proposed superstatistics formalism [C. Beck and E.G.D. Cohen, Physica A **322**, 267 (2003)], used to interpret unconventional statistics. Their interpretation is that unconventional statistics in dynamical systems arise as weighted averages of the ordinary statistics obeyed by these systems over a statistical distribution of background configurations due to fluctuations intrinsic to the background. In this paper we suggest that the same picture can arise because of the intrinsic dynamics of the system. The dynamics of the system and the background, hence, concur together to determine the overall final statistics: differently evolving systems embedded within the same background can yield different statistical distributions. Some simple examples are provided; among them a toy model able to yield a power-law distribution. Also, some recent independent results are quoted, that appear to support this viewpoint.*

PACS: 05.20.-y; 05.40.-a; 05.90.+m

Keywords: statistical physics; superstatistics; turbulence; fluid mechanics

Since several years an active field of study is the investigation of systems where the Probability Density Distribution (PDF) for fluctuations of some relevant quantity is not distributed according to a Gaussian curve. Very often, instead, empirical PDFs from fairly different systems are found to be matched, at least asymptotically and approximately, by power-law curves.

It is tempting to search for a unifying principle for this general feature. Within this context, Tsallis' non-extensive entropy [1] has enjoyed a widespread interest. Tsallis' is a first-principle theory, in that it starts by postulating the existence of a universal functional of the microscopic statistical distribution of the system (Tsallis' generalized entropy), whose analytical form is the same whatever the system studied. The PDF is derived through standard constrained extremization. The only free parameter of the theory, which fully characterizes the system at hand, is the non-extensivity exponent $q$.

Quite recently, Beck [2,3] and Beck and Cohen [4], prompted by the works by Wilk and Wlodarczyk [5], have suggested an alternative approach, able to model non-gaussian PDFs, dubbed *superstatistics*. Let us consider a complex dynamical system $\Sigma$ interacting with a background, in thermodynamical equilibrium with it. The PDF for a relevant thermodynamical variable of the system (say, generalized energy "$E$") obeys the classical Maxwell-Boltzmann distribution[#]

$$P(E) = \beta \exp(-\beta E) \qquad (1)$$

controlled by the parameter (Lagrange multiplier, generalized inverse temperature) $\beta$, which quantifies the influence of the background. The generalized energy E is often the kinetic energy for a structureless particle, but can also be associated to some internal degree of freedom. For the purposes of the present paper, this latter view is more interesting.

Let us now replace this simple background with a more sophisticated one, where the parameter $\beta$ is no longer fixed, but varying: in its turn, it is a statistical quantity, endowed with its own PDF. In the course of its evolution, the system $\Sigma$ interacts with the varying background. It is supposed that the background varies slowly, so that the system has time enough to reach local thermodynamical equilibrium. Any measure, performed over times much longer that those typical of the fluctuations of the background will yield an effective PDF for the system that is the weighted average of (1) over the different realizations of the background:

$$\tilde{P}(E) = \int P(E) f(\beta) d\beta \qquad , \qquad (2)$$

where $f(\beta)$ is the PDF for $\beta$.

By appropriately selecting the PDF $f(\beta)$, power-law PDFs as well as other analytical functions for $\tilde{P}(E)$ can be realized [3,6]. By example, standard Maxwell-Boltzmann statistics is recovered when $f(\beta) \propto \delta(\beta - \beta_0)$. Hence, the superstatistics formalism includes power-law's as well as other non-conventional statistics. Indeed, despite the large interest towards power-law statistics (there are theoretical reasons to expect these distributions to rank in importance with normal distributions-see, e.g., chaps. 4 and 14 in [7]), these make just a fraction of all possible and experimentally realized cases. It even happens that some empirical PDFs, once thought to be fitted by power-law curves, are ascribed to other analytical curves after more sophisticated analysis or repeated measurements (later experiments of Jung and Swinney, quoted in ref. [4], [8]; also, it is known that power laws and other analytical curves such as log-normal can be mistaken in presence of noise and/or small sampling intervals, see chap. 4 in [7]).

---

[#] In this work we will use interchangeably the terms "Maxwell-Boltzmann distribution" and "Gaussian distribution". More properly, when the former applies to the $E$ variable, the second holds for $v = E^{1/2}$ (plus a suitable change in the normalization factor).

In a nutshell, therefore, superstatistics just relies on a tool well known in statistics, that of superposition of distributions-which in the end can be related to Bayes' calculus of probabilities-together with an ansatz about the form for $P(E)$ and a physical interpretation for the other distribution, $f(\beta)$. It is within this last term where all the physics stays.

Superstatistics-in the present formulation-replaces one question ("where do non-gaussian PDFs come from", or "what is the origin of Tsallis' entropy?") by another ("why/how does a dynamical process like turbulence lead to a quasi-steady state in which local control parameter $\beta$ is distributed in a particular way?"). As long as superstatistics postulates that $\beta$ must be a fluctuating stochastic variable, the emphasis is on writing down the PDF $f(\beta)$. Since usually one has information on the system $\Sigma$ but much less on the background, $f(\beta)$ can hardly be deduced from first principles; instead, it is devised *a posteriori*, on the basis of the sought $\widetilde{P}(E)$. Of course, this approach is not completely satisfactory since contains an amount of arbitrariness: a better theory should be able to yield a justification for the appearance of $f(\beta)$ statistics. As an example, one may compare the two papers [2,8]. Both deal with the problem of recovering $\widetilde{P}(E)$ PDFs of velocity differences in a turbulent fluid. *A fortiori*, hence, these papers must determine $f(\beta)$, and they start from the same premises for getting a theoretical picture of the fluctuating background. At a second stage, a very simple different guess (actually, it is slightly more than a different normalization), however, makes the two works reach far different conclusions about the functional form of PDFs $f(\beta), \widetilde{P}(E)$. We stress that both choices are equally likely and arbitrary, since there is not information available to decide between them.

The aim of this work is to suggest a different (not alternative) physical interpretation of superstatistics formalism. It will not be free of the above arbitrariness; that is, some hypotheses must still be imposed. However, these hypotheses appear rather intuitive and-above all-their empirical validation through experiments appears feasible (more about this in the conclusions).

The starting point is noting that Eq. (2) requires ergodicity: in fact, experimentally, one does not measure the integral (2), rather performs an integration over time, whose result is a time average of expressions of kind (1), each weighted by the time the system spends interacting with the background in the state parameterized by β. It translates into an average over the PDF $f(β)$ (Eq. 2) under the hypothesis of ergodicity. In other terms,

$$\tilde{P}_{exp}(E) = \int f_t(t)P(E, \beta(t))dt \ , \tag{3}$$

where $f_t(t)dt$ is the fraction of time the background spends in the state $β(t)$. Under the hypothesis of ergodicity $f_t(t)dt = f(\beta)d\beta$, which leads to Eq. (2). This equality may arise within several different scenarios. We think of three limiting cases. The first is, when one can think just to temporal fluctuations of the variable β (temporal chaos in the background). A second case is when the background is stationary but spatially chaotic. In both cases, even a very regular trajectory leads Σ to sample different configurations of the background. Beck's work emphasizes these two possibilities (see [3]). As a third case, we can think to an erratic dynamics of Σ within a temporally stationary and spatially smooth (but non-uniform) background. In this latter case, $f_t(t)$ is a measure of the time spent by Σ within each β-region as a consequence of its own motion. Of course, in this case, one must also assume that its motion is slow enough (or that β spatially varies slowly enough) for Σ to reach a quasi equilibrium with its surroundings at a given β.

Therefore, under this light, Eq. (3) is reinterpreted as: the effective $\tilde{P}(E)$ comes as an average of $P(E)$ (Eq. 1), weighted by the time spent by Σ close to each region of the background characterized by a (fixed) value β of the control parameter. The point to be stressed is that, now, in eq. (3) $f_t(t)$ is determined by the system's dynamics, not background's. Under this hypothesis, the relation $f_t(t)dt = f(\beta)d\beta$ becomes a definition for $f(β)$.

Of course, intermediate cases can appear, where fluctuating β appears both due to stochasticity intrinsic to the background and in the Σ's motion (see [9]).

In the rest of the paper we will be interested in putting into evidence the importance of $\Sigma$'s dynamics, hence will neglect stochasticity in the background.

Let us illustrate this through a simple toy model: we shall consider an extremely simplified model of a fluid freely flowing into a channel along the $z$ axis. The motion is bounded along a direction, say between $x = 0$ and $x = 1$ (in suitable units), while it unbounded and homogenous along the other ($y$) direction, which is therefore ignorable. Let us establish a temperature gradient between the two boundaries, say a constant gradient. By choosing *ad hoc* units of measure, and neglecting-for easiness-temperature at one boundary with respect to the other one (i.e., $T(x=0) << T(x=1)$), we can write a formal correspondence between local temperature and spatial position:

$$T(x) \equiv x \qquad . \qquad (4)$$

We place ourselves into the Lagrangian viewpoint: we follow a small lump of fluid and compute statistical quantities as averages over the time spent by this particle in a given state.

Let us suppose of regularly measuring the position of the particle and all relevant quantities (in this case, only one: its temperature). Of course, this cannot be done in a continuous fashion but at discrete time steps $\Delta t$. Also, the duration of each measurement be $dt << \Delta t$. For actual computations, we need a rule for time evolution of the particle: let us suppose that its motion be completely random. Instead of postulating complete randomness, one could retain determinism and allow for the fluid flow to approximate an iterated discrete map, in this case a Bernoulli shift map:

$$x(t + \Delta t) = 2x(t) \quad \mathrm{mod}(1) \qquad , \qquad (5)$$

which gives a uniform coverage of the $(0,1)$ interval. For our purposes, it is irrelevant whether the dynamics chosen is realistic or not: we said it is just a toy model.

In both cases, the probability of finding the particle in an interval $dx$ around the point $x$ is simply proportional to $dx$. The overall time spent by the particle within this small interval $dx$, assuming thermal diffusive motion, is given by $dt = dx^2/D_{th}$, with $D_{th}$ local thermal diffusion coefficient. On the basis of Einstein relation, $D_{th} \propto T$ and finally, using Eq. (4), $dt \propto 1/T$. $dt$ must be understood to be small on the macroscopical (transport) scale, but large enough on the microscopical scale for the particle to come to local thermal equilibrium; also, particle's volume is large enough so that fluctuations around the most probable value can be neglected. These are standard constraints in statistical physics. The probability of finding a small volume of fluid with temperature $T$ is, therefore,

$$f(T)dT \propto dt \propto \frac{1}{T}dT \ . \tag{6}$$

By reverting to the inverse generalized temperature $\beta = 1/T = 1/x$ , one gets

$$f(\beta)d\beta \propto \frac{1}{\beta}d\beta \qquad . \tag{7}$$

Any system moving with the fluid and coming into thermal contact with the heat source, has a thermal energy which fluctuates around the local value $1/\beta(x)$, the average energy distribution function sampled is therefore

$$\widetilde{P}(E) \propto \int_{1}^{\infty} \beta \exp(-\beta E)\frac{1}{\beta}d\beta \ . \tag{8}$$

(Notice the extremes of integration in Eq. 8, due to the fact that $0 < T < 1$). The integral (8) is indeed not mathematically rigorous, since $d\beta$ is not a true infinitesimal, but simply a quantity very small at the macroscopical level.

The integral in eq. (8) can be analytically performed and the result is

$$\widetilde{P}(E) \propto \frac{\exp(-E)}{E} \qquad . \tag{9}$$

that is, we have got a PDF that differs from Maxwell-Boltzmann's even though the background is not fluctuating at all: we stress again that the $f(\beta)$ appearing in Eqns. (7,8) is an *effective* PDF as seen on the average by the system, but does not correspond to actual temporal fluctuations of the background, as seen in the laboratory frame.

The departure from Maxwellianity depends both from the background (because of the presence of a spatially varying $\beta$) *and* from the system. We wish to stress again the point raised in the introductory paragraphs: the energy $E$ featuring in Eqns. (8,9) is a measure of the internal energy of the lump of matter, not of its kinetic energy. Indeed, fluctuations in this latter quantity are not determined by the background at all (the equation of motion is given, e.g., in Eq. 5, independent from the background).

Indeed, looking backwards, we can relax the hypotheses done before and allow for fairly different fluid motions: it is not even needed that they be ergodic. Depending upon the kind of motion, different statistics are to be found. By example, if the fluid is at rest, and any fluid particle stays close to its initial location, $x(t) \sim x(0)$, then it would see always one and the same β, thus Maxwell-Boltzmann statistics would be recovered. Another different case arises for laminar motion, when we can approximate the particle motion as ballistic, hence the following chain holds:

$$dt \propto dx \propto dT \rightarrow f(T)dT \propto dt \propto dT \rightarrow f(T) = \text{const} \rightarrow f(\beta)d\beta \propto 1/\beta^2 d\beta$$

This prompts the suggestion that different kinds of motion-embedded in the same background-would lead to different statistics: for example, what if we maintain ergodicity of the motion but allow for a non-uniform spatial distribution?. Let us replace the Bernoulli shift map with another chaotic map, say the logistic map (again, we do not bother about the realism of our choice). This map has an invariant measure which is not constant over $x$: the probability density for finding a point at the location $x$ is

$$p_{LM}(x) = \frac{1}{\pi\sqrt{x(1-x)}} \qquad . \qquad (10)$$

We repeat the same calculations as before, and the temperature PDF becomes

$$f(T) \approx p_{LM}(x)\frac{1}{x} \propto \frac{1}{x^{3/2}\sqrt{1-x}} \qquad . \qquad (11)$$

Notice that Eq. (11) is not normalizable, but it does not lead to any problem: we could circumvent the problem by defining $T = x + \varepsilon$ and then letting $\varepsilon \rightarrow 0$ at the end of the calculation. Instead, we will consider formally as well defined Eq. (11): it gives meaningful final results. We get, as before

$$f(\beta) \propto \frac{1}{\sqrt{(\beta-1)}} \qquad . \qquad (12)$$

which, when inserted into Eq. (2) yields

$$\tilde{P}(E) \propto \frac{\exp(-E)}{E^{3/2}}(1+2E) \qquad (13)$$

In Fig. 1 we plot together the resulting $\tilde{P}(E)$ for both cases (9, 13). The difference between them and from a pure exponential is small but perceivable.

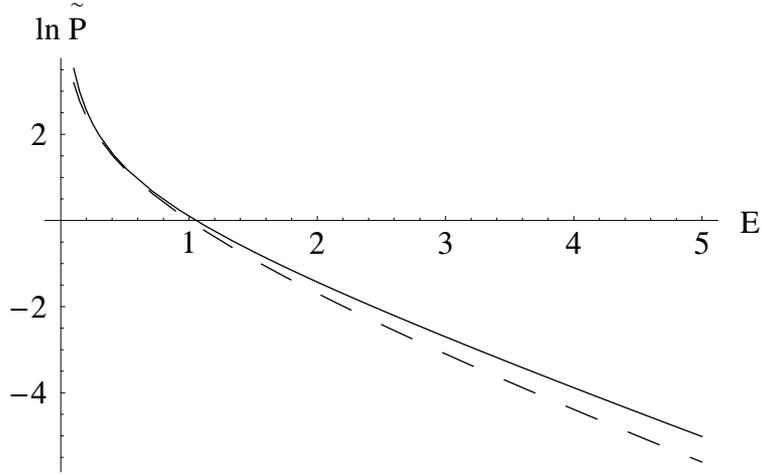

Fig.1: natural logarithm of $\widetilde{P}(E)$ *versus* $E$. Dashed line, $\widetilde{P}(E)$ from Eq. (9); solid line, $\widetilde{P}(E)$ from Eq. (13). A Maxwell-Boltzmann distribution is a pure exponential and would appear as a straight line in this plot. Note that the two curves have been shifted so as to overlap, i.e., they are not normalized.

The fact that both Eq. (9) and (13) are not proper probability distributions, since they are not normalizable, is not really an issue; rather, is due to the fact that the hypotheses the models are built on, break down for very small $E$. A more accurate modelization would impose additional constraints, in the form-by example-of infrared cut-off, such that (9,13) become normalizable.

Apart from this, the two examples we have worked out are very intuitive, but are not the best possible, since the resulting PDFs $\widetilde{P}(E)$ are very close to unperturbed Maxwell-Boltzmann's. The underlying model is very rough and far from any claim of realism. The next example shows that a small modification is sufficient to yield power-law PDFs, which can be of more interest. Let us consider almost the same system as before: the fluid is now bounded between $x = 1$ and $x = L$ (whose value is irrelevant). This time, the temperature (always in suitable units) varies along $x$ as

$$T(x) = \frac{1}{x^{\alpha}}, \quad \alpha > 0 \qquad . \qquad (14)$$

Such a profile can be obtained by the steady state Fourier's law with a thermal conductivity which is a power-law of $x$. Let us assume the fluid motion to be the same as in the first example, i.e., the probability of finding the test fluid particle is simply $f(x)\,dx = dx/L$. From this, straightforwardly,

$$f(T)dT \propto \frac{f(x)}{T}dx \approx \left(\frac{1}{T}\right)^{1+1/\alpha+1} dT \rightarrow f(\beta) \approx \beta^{1+1/\alpha+1-2} \qquad (15)$$

and

$$\widetilde{P}(E) \propto \int_{1}^{L^{\alpha}} \beta \exp(-\beta E)\, \beta^{\frac{1}{\alpha}}\, d\beta \qquad , \qquad (16)$$

that is

$$\widetilde{P}(E) \propto E^{-(1/\alpha+2)} \qquad , \qquad (17)$$

i.e., a power-law PDF. Within Tsallis' framework, result (17) would correspond to a entropic index $q$:

$$\frac{1}{q-1} = \frac{1}{\alpha} + 2 \rightarrow q = 1 + \frac{1}{1/\alpha+2} \qquad . \qquad (18)$$

The goal of this paper was to suggest a different perspective when looking at superstatistics, shifting the emphasis from the spatio-temporal properties of the background to the dynamics of the system. As we have stated earlier, these can be seen as limiting cases of a same picture. Hence, both pictures can realizable in actual situations, and both need the same amount of information to compare with experiment. However, when realized, the situation described here, appears to have some appealing features in terms of simplicity: by example, compare our derivation of Eq. (17) with the path followed in ref. [2] to reach a similar result. Another possible advantage of the present approach is: it has been pointed out that, in order to allow for finite fluctuations in the parameter β, finiteness of the heat reservoir (i.e., a relatively small number of its degrees of freedom) could be necessary. Now, this requirement is not necessary.

It could mentioned that this point of view seems to be supported also by a very recent research [10], appeared while finishing this work: the authors of that work numerically studied a 2D dissipative granular gas, a system well known to feature non-gaussian PDFs. Their findings was that the observed distribution function is governed mainly by the spatial (in)homogeneity of the heating function: Gaussian distributions appearing for uniform heating, and non-Gaussian ones when a non-homogeneous heating profile is imposed. This is, of course, strongly reminiscent of our spatially varying temperature profiles, Eqns. (4,14).

**Acknowledgments**

This work was prompted by comments raised by Prof. M. Wortis. Prof. Beck and the anonymous referees carefully read the manuscript, pointed to some errors and gave useful suggestions. Help from L. Salasnich is acknowledged.